\documentclass[amsthm]{elsart}

\journal{}
\company{}
\newcommand{\xs}{x^\sigma}
\newcommand{\diag}{\mbox{diag}\,}
\newcommand{\mitPsi}{\mathit{\Psi}}

\usepackage{yjsco}
\usepackage{natbib}
\usepackage{amsmath}

\begin{document}
\begin{frontmatter}

\title{The acoustic wave equation in the expanding universe. Sachs--Wolfe theorem}


\author{Wojciech Czaja},
\ead{W.Czaja@oa.uj.edu.pl}
\author{Zdzis{\l}aw A.~Golda},
\ead{zdzislaw.golda@uj.edu.pl}
\author{Andrzej Woszczyna\corauthref{cor}}
\corauth[cor]{Corresponding author.}
\ead{uowoszcz@cyf-kr.edu.pl}

\address{Jagiellonian University, The Faculty of Physics, Astronomy and Applied Computer Science, Astronomical Observatory, 
ul.~Orla 171, 30--244  Krak\'ow}
\address{Copernicus Center for Interdisciplinary Studies,
ul.~Gronostajowa 3,  30--387 Krak\'ow}

\begin{abstract}
In this paper the acoustic field propagating in the early hot ($p=\epsilon/3$) universe of arbitrary space curvature ($K=0, \pm1$) is considered. The field equations are reduced to the d'Alembert equation in an auxiliary static Roberson--Walker space-time. Symbolic computation in {\em Mathematica} is applied.
\end{abstract}
\begin{keyword}
Cosmology; Perturbations, Acoustic waves; Expanding environment; Symbolic computation; {\em Mathematica}
\end{keyword}

\end{frontmatter}

\section{Introduction}

The Sachs--Wolfe theorem (Sachs and Wolfe, 1967, pp. 76--77) contains two separate theses formulated for two different equations of state:  the first ((i) p.~76) for the pressureless matter  ($p=0$), and the second ((ii) p.~77)  for ultrarelativistic gas ($p=\epsilon/3$). The first thesis was recently recalculated and rediscussed by Sachs et al., (2007). In this paper we will concentrate on the second, and for the purposees of this paper we will call it  the {\em acoustic theorem\/} to clearly distinguish between them both.

{\it Acoustic theorem\/} refers to the spatially flat ($K=0$), hot ($p=\epsilon/3$) 
Friedman--Robertson--Walker universe, and to the scalar perturbation propagating in it. It satates, that by the appropriate choice of the perturbation variable one can express the propagation equation in the form of d'Alembert equation in Minkowski space-time.  Scalar perturbations in the flat, early universe propagate in a similar manner to electromagnetic or gravitational waves (Sachs and Wolfe, 1967, p.~79, see also Grishchuk 2007).

On the other hand, the d'Alembert equation describing scalar field in the dust ($p=0$) cosmological model, can be transformed into the d'Alembert equation in the static Robertson--Walker space-time regardless of the universe space curvature (Klainerman and Sarnak, 1981). We may, therefore, suppose that the flatness assumption in the Sachs--Wolfe theorem is excessive and that the theorem is true in the general case. 

Indeed, as we show below,  the scalar perturbations reveal similar features in both, the flat and curved universes, i.e. they propagate as waves in the static Robertson--Walker space-time. The proof of this fact is the subject of this paper. The Sachs--Wolfe {\em acoustic theorem\/} may apply to the microwave background radiation (Sachs et al., 2007).

The paper consists of a brief description of the problem, and the short code in {\em Mathematica}. Throughout this paper  $c = 1$ and  $8\pi G=1$.

\section{Scalar perturbations on the Friedman--Robertson--Walker background}

We adopt Robertson--Walker metrics in spherical coordinates $\xs=\{\eta,\chi,\vartheta,\varphi\}$
\begin{equation}
g_ {(\mathcal {R}\mathcal {W})} = a^2(\eta )\left (
   \begin {array} {cccc}
                      - 1 & 0 & 0 & 0 \\
                     0 & 1 & 0 & 0 \\
                   0 & 0 & \frac {\sin ^2\left (\sqrt {K} \chi 
                   \right)} {K} & 0 \\
            0 & 0 & 0 & \frac {\sin ^2\left (\sqrt {K} \chi 
           \right) \sin ^2 (\vartheta )} {K}
      \end {array}
     \right)
\label{metryka_RW}
\end{equation}
with the scale factor $a(\eta$) appropriate for  the equation of state $p = \epsilon/3$
\begin{equation}
a(\eta) = \frac {\sin  \left (\sqrt {K} \eta \right)} {\sqrt {K}}.
\label{czynnik_skali}
\end{equation}
Consider the scalar perturbations in the synchronous system of reference. The metric tensor correction is 
then determined by two scalar functions  $\mathcal{E}$  and $\mathcal{C}$ (Stewart,~1990)
\begin{eqnarray}
\delta g_{\mu 0}&=&0,\\
\delta g_{m n}&=&a^2\left(\nabla_{\!m}\nabla_{\!n} \mathcal{E}+\frac{1}{3} (\mathcal{C}-\triangle \mathcal{E})g_{m n}\right).
\label{perturbacja_metryki}
\end{eqnarray}
$\triangle$ stands for the Beltrami--Laplace operator on the $\eta=\mbox{const}$ hypersurface. Eventually, 
the metrics takes the form
\begin{equation}
g_ {\mu\nu} =g_ {({\mathcal {R}\mathcal {W})}\mu\nu} +\delta g_{\mu\nu}
\label{perturbacja_metryki_RW}
\end{equation}
In the first order of the perturbation expansion the Einstein equations with the metric tensor (\ref{perturbacja_metryki_RW}) reduce to
\begin{eqnarray}
\label{lambda_RW}
\frac{\partial^2}{\partial\eta^2}\lambda({\xs})
&=&-2\frac{a'(\eta )}{a(\eta )}
\frac{\partial}{\partial\eta}\lambda({\xs})-
\frac{1}{3}\triangle
\left[
\lambda({\xs}) +\mu({\xs})
\right],\\
\frac{\partial^2}{\partial\eta^2}\mu({\xs})
&=&-\left[
2+3c^2(\eta)
\right]\frac{a'(\eta )}{a(\eta )}
\frac{\partial}{\partial\eta}\mu({\xs})+
\left[
\frac13+c^2(\eta)
\right]\times\nonumber
\\
&&
~~~~~~~~~~~~~~~~~~~~~~~~~~~~~~{}\times\left[
\left(
3K+\triangle
\right)\left(\lambda(\xs)
+
\mu(\xs)\right)
\right],
\label{mu_RW}
\end{eqnarray}
where  we put  $\mu = \mathcal{C}$, $\lambda = -\triangle\mathcal{E}$  (to fit to (Lifshitz and Khalatnikov, 1963, Landau and Lifshitz, 2000)). $c(\eta)$ stands for sound velocity. The energy density contrast $\delta=\delta\epsilon/\epsilon$ reads
\begin{equation}
\delta(\xs) = \frac {1} {3\epsilon(\eta) a^2(\eta)}
\left[
3 \frac{a'(\eta )}{a(\eta )} 
\frac{\partial}{\partial\eta}\mu({\xs})
-\left(3K+\triangle\right)\left[\lambda(\xs)+\mu(\xs)\right]\right].
\label{kontrast}
\end{equation}


Let us define a new perturbation variable $\mitPsi$ with help of the second order differential transformation of the density contrast $\delta$
\begin{equation}
\mitPsi(\xs) = \frac {1} {\cos(\sqrt{K}\eta)}
\frac{\partial}{\partial\eta}
\left[
\frac {K} {\tan^2(\sqrt{K}\eta)}
\frac{\partial}{\partial\eta}
\left(
\frac {\tan^2(\sqrt{K}\eta)}{K}\cos(\sqrt{K}\eta)\delta(\xs)
\right)
\right].
\label{zmienna_Psi}
\end{equation}
The function  $\mitPsi(\xs)$ is the solution of the d'Alembert equation
\begin{equation}
\frac{\partial^2}{\partial\eta^2}\mitPsi(\xs)- \frac {1}{3}
\triangle\mitPsi(\xs)=0
\label{rownanie_falowe}
\end{equation}
with the Beltrami--Laplace operator $\triangle={}^{(3)}\!g_{m n}\nabla^m\nabla^n$  acting in the space
\begin{equation}
{}^{(3)\!}g = \left (
   \begin {array} {ccc}
                      1 & 0 & 0 \\
                   0 & \frac {\sin ^2\left (\sqrt {K} \chi 
                   \right)} {K} & 0 \\
            0 & 0 & \frac {\sin ^2\left (\sqrt {K} \chi 
           \right) \sin ^2 (\vartheta )} {K}
      \end {array}
     \right).
\label{operator_Beltrami_Laplacea}
\end{equation}

\begin{thm}[Sachs--Wolfe acoustic theorem for $K=0, \pm 1$]
Scalar perturbations in the hot ($p=\epsilon/3$) Friedman--Robertson--Walker universe of arbitrary space curvature ($K=0, \pm 1$), expressed in terms of  the perturbation variable $\mitPsi$ (\ref{zmienna_Psi}) obey  the wave equation (\ref{rownanie_falowe}) in the static Robertson--Walker space-time $g=\diag(-1,{}^{(3)\!}g)$.
\end{thm}

\begin{pf}

We perform this calculation in {\em Mathematica}. The code is enclosed below, and also can be downloaded from:\\ 
(* http://drac.oa.uj.edu.pl/usr/woszcz/kody/acoucticRWcode.nb *)

\bigskip

\leftline{\indent(* Coordinates *)}
\smallskip

\noindent\({\text{Clear}[K,a,c,\epsilon ,H,\text{Lap},\Psi ]}\)

\noindent\({X=\{\eta ,\chi ,\vartheta ,\varphi \};\text{XX}=\{\eta \_,\chi \_,\vartheta \_,\varphi \_\};} \\
{x=\text{Sequence}\text{@@}X;\text{xx}=\text{Sequence}\text{@@}\text{XX};}\)

\medskip 
\leftline{\indent(* Friedman--Robertson--Walker background for $\epsilon =3p$ *)}
\smallskip

\noindent\({c[\eta \_]=\sqrt{\frac{1}{3}}; a[\eta \_]= \text{Limit}\left[\frac{ \text{Sin}\left[\sqrt{\kappa } \eta \right]}{ \sqrt{\kappa }},\kappa
\to K\right];}\)

\noindent\({H[\eta \_]=\frac{a'[\eta ]}{a[\eta ]^2}\text{//}\text{Simplify}; \epsilon [\eta \_]=\frac{3 K}{a[\eta ]^2}+3 H[\eta ]^2\text{//}\text{Simplify};}\)

\medskip 
\leftline{\indent(* Lifshitz--Khalatnikov perturbation equations *)}
\smallskip

\noindent\({{\lambda ^{(2,0,0,0)}[\text{xx}]=-\frac{2 a'[\eta ] }{a[\eta ]}\partial _{\eta }\lambda [x]-\frac{1}{3} (\text{Lap}[\lambda ][x]+\text{Lap}[\mu
][x]);}}\\
{{\mu ^{(2,0,0,0)}[\text{xx}]=-\left(2+3 c[\eta ]^2\right) \frac{ a'[\eta ] }{a[\eta ]}\text{  }\partial _{\eta }\mu [x]+\frac{1}{3}\left(1+3
c[\eta ]^2\right)}}\)

\noindent~~~~~~~~~~~~~~~~~~~~~~~~~~~~~~~~~~~~~~~~~~~~~~~~~~~~~~~~~~~~~~~
 \(
((\text{Lap}[\lambda ][x]+3 K \lambda [x])+(\text{Lap}[\mu ][x]+3 K \mu [x]));
\)
\leftline{}

\medskip 
\leftline{\indent(* The density contrast *)}
\smallskip

\noindent\({{\delta [\text{xx}]}{=}{ }{\frac{1 }{3 \epsilon [\eta ]a[\eta ]^2}\left( -((\text{Lap}[\lambda ][x]+3 K \lambda [x])+(\text{Lap}[\mu
][x]+3 K \mu [x]))+3\frac{a'[\eta ]}{a[\eta ]} \partial _{\{\eta ,1\}}\mu [x]\right)}}\)

\noindent~~~~~~~~~~~~~~~~~~~~~~~~~~~~~~~~~~~~~~~~~~~~~~~~~~~~~~~~
~~~~~~~~~~~~~~~~~~~~~~~~~~~~~~~~~~~~~~~~~~~\text{//}\text{Simplify};

\medskip 
\leftline{\indent(* Higher order derivatives *)}
\smallskip

\noindent\({\lambda ^{(\text{k$\_$Integer}\text{/;}k>2,0,0,0)}[\text{xx}]\text{:=}D\left[\lambda ^{(2,0,0,0)}[x],\{\eta ,k-2\}\right]}\\
{\mu ^{(\text{k$\_$Integer}\text{/;}k>2,0,0,0)}[\text{xx}]\text{:=}D\left[\mu ^{(2,0,0,0)}[x],\{\eta ,k-2\}\right]}\\
{\lambda ^{(2,\text{d1$\_$},\text{d2$\_$},\text{d3$\_$})}[\text{xx}]=D\left[\lambda ^{(2,0,0,0)}[x],\{\chi ,\text{d1}\},\{\vartheta ,\text{d2}\},\{\varphi
,\text{d3}\}\right];}\\
{\mu ^{(2,\text{d1$\_$},\text{d2$\_$},\text{d3$\_$})}[\text{xx}]=D\left[\mu ^{(2,0,0,0)}[x],\{\chi ,\text{d1}\},\{\vartheta ,\text{d2}\},\{\varphi
,\text{d3}\}\right];}\)

\medskip 
\leftline{\indent(* Laplacian in the maximally symmetric 3-$\dim$ curved space *)} 
\smallskip

\noindent\(
{
w[\text{K$\_$},\chi \_]=\text{Limit}\left[2\sqrt{\kappa} \text{Cot}\left[\sqrt{\kappa } \chi \right],\kappa \to K\right];
}\\
{\text{Lap}[\text{f$\_$}][\text{xx}]=\frac{w[K,\chi ]^2}{4\text{Cos}\left[\sqrt{K} \chi \right]^2} \left(\text{Csc}[\vartheta ]^2 \partial _{\{\varphi
,2\}}f[x]+\text{Cot}[\vartheta] \partial _{\vartheta }f[x]
+\partial _{\{\vartheta ,2\}}f[x]\right)}\\
{~~~~~~~~~~~~~~~~~~~~~~~~~~~~~~~~~~~~~~~~~~~~~~~~~~~~~~~~~~~~~~{}+w[K,\chi ] \partial_{\chi }f[x]+ \partial_{\{\chi
,2\}}f[x];}\)

\medskip 
\leftline{\indent(* The gauge-invariant variable $\Psi$ *)}
\smallskip

\noindent\({\tau [\text{K$\_$},\eta \_]=\text{Limit}\left[\frac{ \text{Tan}\left[\sqrt{\kappa } \eta \right]^2}{\kappa  },\kappa \to K\right];}\\
{ \Psi [\text{xx}]=\left(\frac{1}{\text{Cos}\left[\sqrt{K} \eta \right]}\partial _{\{\eta ,1\}}\left(\frac{1}{\tau [K,\eta ]} \partial _{\eta
}\left(\tau [K,\eta ] \text{Cos}\left[\sqrt{K} \eta \right] \delta [x]\right)\right)\right);}\)

\medskip
\leftline{\indent(* The wave equation *)}
\smallskip

\noindent\({{ \partial _{\{\eta ,2\}}\Psi [x]-\frac{1}{3}\text{Lap}[\Psi ][x]\text{==}0;}}\)
\medskip 

\noindent\({\text{Timing}[\text{Simplify}[\%,\text{TimeConstraint}\to 5000]]}\)

\end{pf}

\section{Discussion}

By obtaining equation (\ref{rownanie_falowe}) we have expressed the problem of cosmological density perturbations  in terms of the {\em semiclassical theory} (the field theory in curved space-time --- see Birrell and Davies, 1982). Consequently, the key facts resulting from the {\em semiclassical theory} apply also to the perturbed expanding universe.

Scalar perturbations in the early universe form the acoustic field. Acoustic waves are dispersed by the space curvature. The dispersion relation for the equation (\ref{rownanie_falowe}) takes the form (Birrell and Davies, 1982, eq.~(5.27))
\begin{equation}
\omega=\sqrt{\frac{1}{3} (k^2-K)}
\label{relacja_dyspersyjna}
\end{equation}
and the group velocity reads
\begin{equation}
v_{\rm g}=\frac{\partial}{\partial k}\omega=\frac{k}{\sqrt{3}\sqrt{k^2-K}}.
\label{predkosc_grupowa}
\end{equation}
In flat space ($K=0$) the group velocity is constant and equal to  $1/\sqrt3$. This fact has been already established by Sachs and Wolfe (Sachs and Wolfe, 1967). 

Beyond the $K=0$ case the dispersion relation is nonlinear. In the space of negative curvature $K=-1$ the waves behave as the scalar  field with the mass $m=1$.  The group velocity is the function of $k$ and decreases to zero in the $k\to0$ limit
\begin{equation}
\lim_{k\to0}v_{\rm g}=0
\label{granica_predkosci_grupowej}
\end{equation}
while the limit frequency is still positive 
\begin{equation}
\lim_{k\to0}\omega=\frac{1}{\sqrt3}.
\label{granica_czestosci}
\end{equation}
Acoustic behaviour does not extend to the solutions of imaginary $k$: $k^2\in\{-1,0\}$ (supplementary series (Gelfand et al., 1966, Neimark, 1964), supercurvature modes (Lyth and Woszczyna, 1995)). 

In the closed universe ($K=1$) the wave numbers are integer and $1<k$ . The general solution is a countable combination of the hyperspherical functions.

On basis of $\mitPsi(\xs)$ one can reconstruct the solution for the density contrast $\delta(\xs)$ by means of the reciprocal relation (\ref{zmienna_Psi}). (Compare analogue procedure for scalar field (Klainerman and Sarnak, 1981). By reconstructing $\delta(\xs)$ one is forced to restore the undesirable phenomenon of gauge ambiguity.


\end{document}